\documentclass[onecolumn,secnumarabic,amssymb, amsmath, nofootinbib,tightenlines,
nobibnotes, aps, prl,epsfig]{revtex4}
\usepackage{graphicx}
\usepackage{dcolumn}
\usepackage{bm}
\begin{document}
\preprint{APS/123-QED}
\title{Color dipole cross-section from unifying the color dipole
picture and improved saturation models}

\author{G.R.Boroun}%
 \email[]{boroun@razi.ac.ir }
\affiliation{ Department of Physics, Razi University, Kermanshah
67149, Iran}


\date{\today}
\begin{abstract}
We present an analysis of the color-dipole picture (CDP) for
determination of the gluon density at low-$x$ which is obtained
from the Altarelli-Martinelli equation by expansion at distinct
points of expansion. The dipole cross-sections with respect to the
improved saturation model of Bartels-Golec-Biernat-Kowalski (BGK)
are obtained in a wide range of transverse sizes $r$ and compared
with the Golec-Biernat-W$\ddot{\mathrm{u}}$sthoff (GBW) model. We
find that the model gives a good description of the dipole
cross-section at large $r$ which confirms saturation and matches
the perturbative QCD result at a small $r$ due to the significant
role of the running of the gluon distribution. The transition
between these regions occurs with decreasing transverse sizes with
a decrease of Bjorken $x$ and dependence on the expansion point.\\

\end{abstract}
 \pacs{***}
\maketitle
\subsection{I. Introduction}
One of the remaining challenges in particle and nuclear physics at
low $x$ in future colliders is understanding the structure of
hadrons in terms of gluons. These novel opportunities will be
opened at new-generation facilities, such as the large hadron
electron collider (LHeC) [1], the Future Circular hadron-electron
Colliders (FCC-he)[2] and the Electron-Ion Collider (EIC) [3]. The
electron-proton center-of-mass energies in the LHeC and FCC-he are
proposed to be $\sqrt{s}=1.3~\mathrm{TeV}$ and
$\sqrt{s}=3.5~\mathrm{TeV}$ respectively. The kinematics of the
LHeC in the $(x,Q^2)$ plane in neutral currents reaches $\simeq
1~\mathrm{TeV}^2$ and $\simeq 10^{-6}$ for $Q^2$ and $x$
respectively [1,4]. The center-of-mass energies in electron-ion
colliders (i.e., EIC and EIcC) are 15-20 $\mathrm{GeV}$ for EIcC
and 30-140 $\mathrm{GeV}$ for EIC [3].\\
Gluons, which mediate the strong interaction, contain essential
information about the hadron and play a crucial role in the
properties of those. The gluon distribution in deep inelastic
scattering (DIS) has been studied by authors in Refs.[5-8]. The
gluon density is dominant at low $x$ ( where
$x{\cong}\frac{Q^2}{W^2}{\ll}1$ and $W$ denotes the
virtual-photon-proton center of mass energy) and this
determination could be a test of perturbative QCD (pQCD) or a
probe of new effects. One of the interesting models is the color
dipole model (CDM), which was first proposed by the authors in
Refs.[9-10] and then extended in Refs.[11-17]. This model is
represented by the hadronic fluctuations (in terms of the
$q\overline{q}$ ) that interact in a gauge-invariant manner as
color-dipole states with the proton via two-gluon exchange. In
CDM, the gluon saturation effects are important at low scales.
Indeed, the growth of the gluon density is tamed, which is related
to unitarity. This result from gluon recombination was defined in
terms of the nonlinear evolution equations [18, 19]. A further
refinement of the saturation was proposed by the
Balitsky-Kovchegov (BK) [20, 21] equation. In this model, the
dipole scattering amplitude was proposed in terms of the Wilson
operators.\\
The dipole cross-section is directly connected via a Fourier
transform to the unintegrated gluon distribution (UGD) at small
$x$, whose evolution in $x$ is regulated by the Balitsky-
Fadin-Kuraev-Lipatov (BFKL) equation [22]. The BFKL equation
governs the evolution of the UGD,
 where the $k_{t}$-factorization is used in the high energy limit in which the QCD
interaction is described in terms of the quantity which depends on
the transverse momentum of the gluon (i.e., $k^{2}_{t}$) [23].
Interpretation of the high-energy interactions predicts that the
small $x$ gluons in a hadron wavefunction should form a Color
Glass Condensate [24] which describes the over-populated gluonic
state.\\
Dipole representation provides a convenient description of DIS at
small $x$, where the dipole cross-section of the scattering
between the virtual photon $\gamma^{*}$ and the proton is related
to the imaginary part of the $(q\overline{q})p$ forward scattering
amplitude. The dipole cross-section in the Golec-Biernat and
W$\ddot{\mathrm{u}}$sthoff (GBW) [10, 25] model is a good
description of the DIS data with only three fitted parameters
((i.e.,$\sigma_{0}$, $x_{0}$ and $\lambda$ ) ), and takes the
following form
\begin{eqnarray}
\sigma_{\mathrm{dip}}(x,\mathbf{r})=\sigma_{0}\bigg{\{}1-
\exp\bigg{(}-r^2Q_{\mathrm{s}}^2(x)/4\bigg{)} \bigg{\}},
\end{eqnarray}
where $Q_{\mathrm{s}}(x)$ plays the role of the saturation
momentum, parametrized as
$Q_{\mathrm{s}}^2(x)=Q_{0}^{2}(x_{0}/x)^\lambda$ (with
$Q_{0}^{2}=1\mathrm{GeV}^2$) and $r$ is the transverse dipole size
of the $q\overline{q}$ pair. The dipole cross-section saturates to
a constant value at large $r$,
$\sigma_{\mathrm{dip}}{\simeq}\sigma_{0}$, and vanishes for small
$r$ which makes the color transparency,
$\sigma_{\mathrm{dip}}{\sim}r^2$, which is a purely pQCD
phenomenon. As is well known, this model is reasonable only at
small transverse momenta $k_{t}$ (after the Fourier transform), as
its exponential decay contradicts the expected perturbative
behaviour at large $k_{t}$. The Bartels-Golec-Biernat-Kowalski
(BGK) model [26], is another phenomenological approach to dipole
cross-section and reads
\begin{eqnarray}
\sigma_{\mathrm{dip}}(x,\mathbf{r})=\sigma_{0}\bigg{\{}1-
\exp\bigg{(}
\frac{-\pi^2r^2\alpha_{s}(\mu_{r}^2)xg(x,\mu_{r}^2)}{3\sigma_{0}}
\bigg{)} \bigg{\}}.
\end{eqnarray}
The evolution scale $\mu_{r}^2$ is connected to the size of the
dipole by $\mu_{r}^2=\frac{C}{r^2}+\mu^2_{0}$ and
$xg(x,\mu_{r}^2)({\equiv}G(x,\mu_{r}^2))$ is the gluon
distribution function. The parameters $C$ and $\mu_{0}$ are
determined [25] from the fits to the HERA data and obtained owing
to the dipole quark mass. The Bjorken variable $x$ is also
modified, since the photon wave function depends on the mass of
the quark in the $q\overline{q}$ dipole, by the following form
\begin{eqnarray}
x{\rightarrow}\overline{x}_{f}=x\bigg{(}1+\frac{4m_{f}^{2}}{\mu_{r}^2}\bigg{)}
=\frac{\mu_{r}^2+4m^2_{f}}{\mu_{r}^2+W^2},
\end{eqnarray}
Therefore the coefficients are defined [25] according to the quark
masses $m_{f}=0.14, 1.4$ and $4.6~\mathrm{GeV}$ (for light, charm
and bottom respectively) in Table I.\\
In small dipole size, $\sigma_{\mathrm{dip}}(x,\mathbf{r})$ (i.e.,
Eq.(2)) is proportional to the gluon distribution from the DGLAP
evolution as
\begin{eqnarray}
\sigma_{\mathrm{dip}}(x,\mathbf{r}){\simeq}
\frac{\pi^{2}r^2}{3}\alpha_{s}(C/{r}^2)xg(x,C/{r}^2),
\end{eqnarray}
and has the property of color transparency. For large dipole size,
Eq.(2) in the limit $\mu_{r}^2{\simeq}\mu_{0}^{2}$ reads
\begin{eqnarray}
\sigma_{\mathrm{dip}}(x,\mathbf{r}){\simeq}\sigma_{0}\bigg{\{}1-
\exp\bigg{(}
\frac{-\pi^2r^2\alpha_{s}(\mu_{0}^2)xg(x,\mu_{0}^2)}{3\sigma_{0}}
\bigg{)} \bigg{\}},
\end{eqnarray}
where in this limit, the saturation scale of the GBW model is
proportional to the gluon distribution at the scale $\mu_{0}^{2}$
by the following form
\begin{eqnarray}
Q_{s}^{2}(x)=\frac{4\pi^{2}}{3\sigma_{0}}\alpha_{s}(\mu_{0}^2)xg(x,\mu_{0}^2).
\end{eqnarray}
The gluon distribution in the BGK model is usually [25-27] evolved
with the DGLAP equations truncated to the gluonic sector and
parametrized at the initial scale of $\mu^{2}_{0}$. The main
purpose of this presentation is to parametrization of the color
dipole cross-section with the gluon distribution from the color
dipole picture. The paper is organized as follows. In Section II
we present the gluon distribution from the color dipole picture (
CDP). In Section III we presents the comparison of the model
results with the GBW model. Finally, we summarize our findings in
Conclusions.\\
\begin{table}[h]
\centering \caption{The coefficients are summarized with respect
to Fits 0-2 in Ref.[25]. }\label{table:table1}
\begin{minipage}{\linewidth}
\renewcommand{\thefootnote}{\thempfootnote}
\centering
\begin{tabular}{|l||c|c|c||c|c|c||c|c|} \hline\noalign{\smallskip} $\mathrm{Fit}$ & $m_{l}$ & $m_{c}$ & $m_{b}$ & $\sigma_{0}[\mathrm{mb}]$ & $\lambda$ & $x_{0}/10^{-4}$ & C & $\mu_{0}^{2}[\mathrm{GeV}^{2}]$   \\
\hline\noalign{\smallskip}
0  & 0.14 & - & - & 23.58${\pm}$0.28& 0.270${\pm}$0.003& 2.24${\pm}$0.16& -& - \\
1  & 0.14 & 1.4 & - & 27.32${\pm}$0.35& 0.248${\pm}$0.002& 0.42${\pm}$0.04& -& - \\
2  & 0.14 & 1.4 & 4.6 & 27.43${\pm}$0.35& 0.248${\pm}$0.002& 0.42${\pm}$0.04& -& - \\
\hline\noalign{\smallskip}
1  & 0 & 1.3 & - & 22.60${\pm}$0.26& -& -& 0.29${\pm}$0.05& 1.85${\pm}$0.20 \\
2  & 0 & 1.3 & 4.6 & 22.93${\pm}$0.27& -& -& 0.27${\pm}$0.04& 1.74${\pm}$0.16 \\
\hline\noalign{\smallskip}
\end{tabular}
\end{minipage}
\end{table}

\subsection{II. The gluon distribution from the color dipole picture}

In QCD, structure functions are defined as convolution of
universal parton momentum distributions inside the proton and
coefficient functions, which contain information about the
boson-parton interaction. The longitudinal structure function is
directly related to the singlet and gluon distributions in the
proton [28], and defined as a convolution integral over
$F_{2}(x,Q^2)$ and the gluon distribution $xg(x,Q^2)$ by  an
effect of order $\alpha_{s}(Q^2)$ as
\begin{eqnarray}
F_{L}(x,Q^2)=C_{\mathrm{L,ns+s}}(\alpha_{s}(Q^2),x){\otimes}F_{2}(x,Q^2)
+<e^2>C_{\mathrm{L,g}}(\alpha_{s}(Q^2),x){\otimes}xg(x,Q^2),
\end{eqnarray}
where $<e^2>=n_{f}^{-1}\sum_{i=1}^{n_{f}}e_{i}^{2}$ and the symbol
$\otimes$ denotes convolution according to the usual prescription.
The standard collinear factorization formula for the longitudinal
structure function at low x reads
\begin{eqnarray}
F_{L}(x,Q^2)=<e^2>C_{\mathrm{L,g}}(\alpha_{s}(Q^2),x){\otimes}xg(x,Q^2),
\end{eqnarray}
where the gluonic coefficient function $C_{\mathrm{L,g}}$ can be
written in a perturbative expansion as follows [29]
\begin{eqnarray}
C_{\mathrm{L,g}}(\alpha_{s}(Q^2),x)=\sum_{n=0}\bigg{(}
\frac{\alpha_{s}}{4\pi} \bigg{)}^{n+1}c_{L,g}^{(n)}(x),
\end{eqnarray}
where $n$ denotes the order in running coupling. \\
Using the expansion method [30] for the gluon distribution
function at an arbitrary point $z=a$ as
\begin{eqnarray}
G(\frac{x}{1-z})|_{z=a}=G(\frac{x}{1-a})+\frac{x}{1-a}(z-a)\frac{{\partial}G(\frac{x}{1-a})}{{\partial}x}+O(z-a)^{2},
\end{eqnarray}
where the series
\begin{equation}
\frac{x}{1-z}|_{z=a}=\frac{x}{1-a}\sum_{k=1}^{\infty}[1+\frac{(z-a)^{k}}{(1-a)^{k}}],
\end{equation}
is convergent for $|z-a|<1$. After doing the integration and
retaining terms only up to the first derivative in the expansion,
we have
\begin{eqnarray}
F_{L}(x,Q^{2})=<e^{2}>A(x,Q^{2})G(\frac{x}{1-a}(1-a+\frac{B(x,Q^{2})}{A(x,Q^{2})}))=<e^{2}>A(x,Q^{2})G(x_{g},Q^2),
\end{eqnarray}
where $x_{g}=kx$, $k=\frac{1}{1-a}(1-a+\frac{B}{A})$ and
$\mathrm{^{"}a^{"}}$ has an arbitrary value $0{\leq}a{<}1$ [31].
In Eq.(12) the kernels $A$ and $B$ read
\begin{eqnarray}
A(x,Q^{2})=\int_{0}^{1-x}\frac{1}{1-z}C_{\mathrm{L,g}}(\alpha_{s},1-z)dz,
\end{eqnarray}
and
\begin{eqnarray}
B(x,Q^{2})=\int_{0}^{1-x}\frac{z-a}{1-z}C_{\mathrm{L,g}}(\alpha_{s},1-z)dz.
\end{eqnarray}
At the leading-order (LO) approximation, Eq.(12) can be rewritten
as
\begin{eqnarray}
F_{L}(x,Q^{2})=\frac{10\alpha_{s}}{27\pi}G(\frac{x}{1-a}(\frac{3}{2}-a)).
\end{eqnarray}
This result reproduced the longitudinal structure function
determined by the authors in Ref.[32] if we expand the gluon
distribution around the point $a=0.666$ as
\begin{eqnarray}
F_{L}(x,Q^{2})=\frac{10\alpha_{s}}{27\pi}G(2.5x,
Q^{2})(=\mathrm{Ref.[32]}).
\end{eqnarray}
For a wide range of different expanding points, the gluon
distribution is defined into the longitudinal structure function
by the following form
\begin{eqnarray}
G(x,Q^2)=\frac{3\pi}{\alpha_{s}(Q^2)\sum_{i}{e_{i}^{2}}}F_{L}(k_{L}x,Q^{2})
\end{eqnarray}
where
\begin{eqnarray}
k_{L}=\frac{2(1-a)}{3-2a}.
\end{eqnarray}
 In terms of the photoabsorption cross-section, $\sigma_{\gamma^{*}_{L}p}$, the longitudinal structure
function is given by
\begin{eqnarray}
F_{L}(W^2(=\frac{Q^2}{x}),Q^2)=\frac{Q^2}{4{\pi^2}\alpha}\sigma_{\gamma^{*}_{L}p}(W^2(=\frac{Q^2}{x}),Q^2),
\end{eqnarray}
where $\sigma_{\gamma^{*}_{L}p}(W^2,Q^2)$ is derived from an
ansatz [9,11-12,14-15] for the $W$-dependent dipole cross-section
by the following form
\begin{eqnarray}
\sigma_{\gamma^*_Lp} (W^2,Q^2)=\frac{\alpha R_{e^+e^-}}{3\pi}
\sigma^{(\infty)} (W^2)I_L (\eta, \mu) G_L (u)
=\frac{\sigma_{\gamma p} (W^2)} {\lim\limits_{\eta \to \mu (W^2)}
I_T \left( \frac{\eta}{\rho}, \frac{\mu (W^2)}{\rho}\right) G_T
(u)} I_L (\eta, \mu) G_L (u)
\end{eqnarray}
where $R_{e^+e^-}=3\sum_{i}{e_{i}^{2}}$ and $\sigma^{(\infty)}
(W^2)$, which stems from the normalization of the $q \bar
q$-dipole proton cross-section, is replaced owing to the  smooth
transition to $Q^2 = 0$ photoproduction [16]. The quantity $I_L
(\eta, \mu)$ is given by
\begin{eqnarray}
 I_L (\eta, \mu)=\frac{\eta - \mu}{\eta}
\times \left( 1 - \frac{\eta}{\sqrt{1+4 (\eta - \mu)}} \right.
\left.  \times \ln \frac{\eta (1+ \sqrt{1+4(\eta - \mu)})}{4
\mu-1-3\eta + \sqrt{(1+4(\eta - \mu))((1+\eta)^2 - 4 \mu)}}
\right),
\end{eqnarray}
where
\begin{eqnarray}
 \eta \equiv \eta (W^2,Q^2) = \frac{Q^2 +
m^2_0}{\Lambda^2_{sat} (W^2)},
\end{eqnarray}
and
\begin{eqnarray}
\mu \equiv \mu(W^2) = \eta (W^2,Q^2 = 0) =
\frac{m^2_0}{\Lambda^2_{sat} (W^2)},
\end{eqnarray}
with the saturation scale $\Lambda^2_{sat}(W^2)$,
\begin{eqnarray}
 \Lambda^2_{sat} (W^2) = C_1 \left(
\frac{W^2}{1 {\rm GeV}^2} \right)^{\LARGE{C_2}},
\end{eqnarray}
and constant parameters ,based on Ref.[16], read
\begin{eqnarray}
m^2_0=0.15~ {\rm GeV}^2, C_1=0.31~ {\rm GeV}^2;~~~C_2 = 0.29.
\end{eqnarray}
The dipole cross-section $\sigma^{(\infty)}(W^2)$, in Eq.(20), is
evaluated from the photoproduction cross-section as [15]
\begin{eqnarray}
\sigma^{(\infty)}(W^2)=\frac{3\pi}{\alpha
R_{e^+e^-}}\frac{\sigma(W^2)}{\ln\frac{\rho}{\mu}},
\end{eqnarray}
where
\begin{eqnarray}
\sigma(W^2)=0.003056\bigg{(}34.71+\frac{0.3894\pi}{M^2}\ln^{2}\frac{W^2}{(M_{p}+M)^2}
\bigg{)}+0.0128\bigg{(} \frac{(M_{p}+M)^2}{W^2} \bigg{)}^{0.462},
\end{eqnarray}
with $\sigma(W^2)$ is given in unit of mb, and
$M=2.15~\mathrm{GeV}$, $M_{P}$ is proton mass in unit of
$\mathrm{GeV}$. The parameter $\rho$ is related to the
longitudinal-to-transverse ratio of the photoabsorption
cross-sections by the constant value $\frac{4}{3}$ owing to
Refs.[15-17]. In the photoproduction limit (i.e., $Q^2=0$), where
$\eta{\rightarrow}\mu(W^2)$,
$G_{T}(u{\equiv}\frac{\xi}{\eta}){\simeq}1$. The parameter
$\xi$~\footnote{The constant parameter $\xi$ restricts the masses
of the contributing mass $q \bar q$ states via $$ M^2_{qq} \le
m^2_1 (W^2) = \xi \Lambda^2_{sat} (W^2). $$} is fixed at $\xi =
\xi_0 = 130$ and therefore
\begin{eqnarray}
\lim_{\eta \to \mu (W^2)} I^{(1)}_T \left( \frac{\eta}{\rho},
\frac{\mu (W^2)}{\rho} \right) = \ln \frac{\rho}{\mu (W^2)}.
\end{eqnarray}
The function $G_{L}(u)$ is given by
\begin{eqnarray}
 G_L (u) = \frac{2u^3 + 6u^2}{2 (1+u)^3} \simeq \left\{
\begin{array}{l@{\quad,\quad}l}
3 \left( \frac{\xi}{\eta}\right)^2 & (\eta \gg \xi),\\
1 - 3 \left( \frac{\eta}{\xi} \right)^2 & (\eta \ll \xi),
\end{array} \right.
\end{eqnarray}
In conclusion, in the scale $\mu_{r}^2$, the gluon distribution in
the color dipole picture is defined by the following form
\begin{eqnarray}
G(x,\mu_{r}^2)=\frac{9\mu_{r}^2}{4\pi \alpha \alpha_{s}(\mu_{r}^2)
R_{e^+e^-} }\frac{\sigma(W^{2*})}{\ln\frac{\rho}{\mu^{*}}}I_L
(\eta^{*}, \mu^{*}) G_L (u^{*}),
\end{eqnarray}
where $W^{2*}=k_{L}^{-1}W_{r}^{2}=k_{L}^{-1}({\mu_{r}^{2}}/{x})$,
$\eta^{*}= ({\mu_{r}^2 + m^2_0})/{\Lambda^2_{sat} (W^{2*})}$ and
$\mu^{*}={m^2_0}/{\Lambda^2_{sat}(W^{2*})}$. The effective dipole
cross-section is evaluated according to the following form
\begin{eqnarray}
\sigma_{\mathrm{dip}}(x,\mathbf{r})=\sigma_{0}\bigg{\{}1-
\exp\bigg{(} -\frac{3\pi(C+\mu_{0}^{2}r^2)}{4\sigma_{0}\alpha
R_{e^+e^-}} \frac{\sigma(W^{2*})}{\ln\frac{\rho}{\mu^{*}}}I_L
(\eta^{*}, \mu^{*}) G_L (u^{*}) \bigg{)} \bigg{\}}
\end{eqnarray}
as a function of $r$ and the expansion point $a$ in different
values of $x$, which also depends on the quark effective mass. The
relation of the CDP gluon density in a proton to the dipole
cross-section in the BGK model will be calculated in the next
section.\\


\subsection{III. Numerical Results}

The saturation scale with the $x$ and $a$ dependencies from the
color dipole models at the scale $\mu_{0}^{2}$ is given by
\begin{eqnarray}
Q_{s}^{2}(x,a)=\frac{3\pi\mu_{0}^2}{\alpha \sigma_{0} R_{e^+e^-}
}\frac{\sigma(W_{0}^{2*})}{\ln\frac{\rho}{\mu_{0}^{*}}}I_L
(\eta_{0}^{*}, \mu_{0}^{*}) G_L (u_{0}^{*}),
\end{eqnarray}
where this saturation scale connects the GBW form of the dipole
cross-section with the CDP. In Fig.1, we compare the saturation
scales  from the expansion points with the GBW model with charm
from the Fits in Table I, as the coefficients $\sigma_{0}$,
$\lambda$ and $x_{0}$ are read from the first three rows of Table
I and the coefficients $C$ and $\mu_{0}^{2}$ are read in the next
two rows. We observe that the results with increases in the
expansion point are close to the GBW model from Fits 0,1 in Table
I. As a result, at $x=10^{-6}$ the saturation scale is order by
$\mathrm{GBW}|_{Q_{s}^{2}{\approx}~2-3~\mathrm{GeV}^2}>\mathrm{Fit~
0,1 }|_{Q_{s}^{2}{\approx}~1-2~\mathrm{GeV}^2}$ and at
$10^{-4}<x<10^{-2}$ the results are equal.\\
In Fig.2, we show the ratio of dipole cross-sections
($\sigma_{\mathrm{dip}}/\sigma_{0}$) in accordance with the active
flavor numbers and the quark mass effects. Quark mass effects are
taken as zero for a massive quark $i$ when $\mu^2<m_{i}^{2}$ and
the quark treated as fully active when $\mu^2>m_{i}^{2}$. The
ratio of dipole cross-sections is obtained, in Figure 2, by
solving Eq.(31) using the CDP $xg(x,\mu_{r}^{2})$ at the expansion
points and compared with the GBW model with the heavy quark
contributions. The results are shown for the selected dipole
transverse size owing to the Fits 0-2.
\begin{figure}
\centerline{
\includegraphics[width=0.58\textwidth]{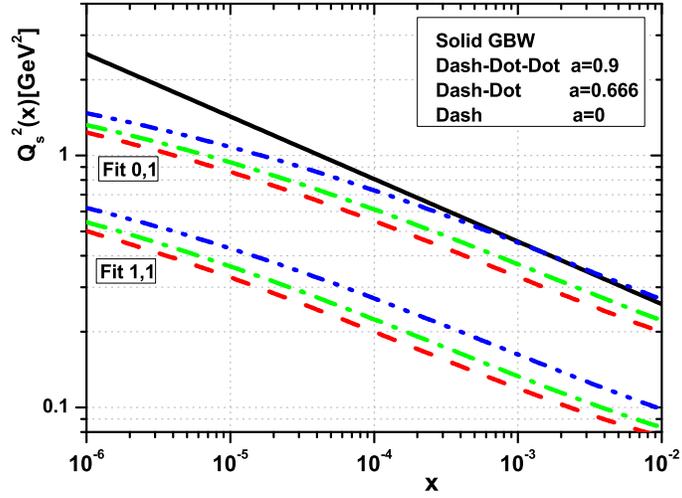}}
\caption{The saturation scale at the scale $\mu_{0}^{2}$ in the
CDP model with the expansion points (a=0, 0.666 and 0.9) with the
parameters from the fits in Table 1 compared with the GBW model
(solid line) with the charm contribution.}\label{Fig1}
\end{figure}
\begin{figure}
\centerline{
\includegraphics[width=0.9\textwidth]{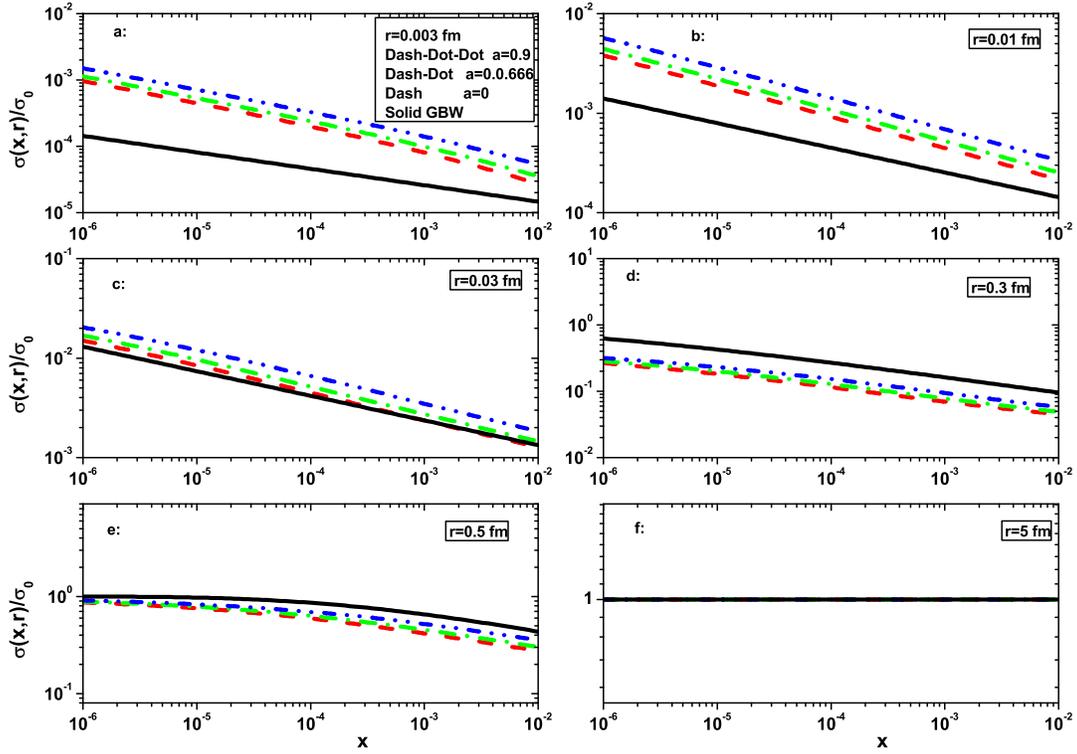}}
\caption{Comparison of the $\sigma_{\mathrm{dip}}(x,r)/\sigma_{0}$
obtained by solving Eq.(31) using the CDP $xg(x,\mu_{r}^{2})$ at
the expansion points with the GBW (solid curves) between
thresholds. (a) r=0.003 fm, Fit 2,2. (b) r=0.01 fm, Fit 2,2. (c)
r=0.03 fm, Fit 1,1. (d) r=0.3 fm, Fit 1,1. (e) r=0.5 fm, Fit 0,1.
(c) r=5 fm, Fit 0,1.}\label{Fig2}
\end{figure}
We observe that our results are sensitive to the expansion point
when we compared with the GBW model with the heavy quarks
contributions. The indicated values of $r$ in Fig.2 are in
accordance with the heavy flavors of fit results in Table I. It is
clear that for large values of $r$ the two functions are very
close, and they differ in the small-$r$ region where the running
of the gluon distribution and its expansion starts to play a
significant role. As a result, the DGLAP improved model with the
gluon distribution in the CDP model can be extending to large
values of $Q^2$ (small dipole sizes).\\
In Fig.3, we have calculated the $r$ dependence, at low $x$
($x=10^{-3}$ and  $x=10^{-6}$), of the ratio
$\sigma_{\mathrm{dip}}(x,r)/\sigma_{0}$ (i.e., Eq.(31)) owing to
the expansion method. Results of calculations with  Fits 1,1 and
2,2 (in Table I) and comparison with the GBW model with the charm
and bottom contributions are presented in Fig.3.
\begin{figure}
\centerline{
\includegraphics[width=0.65\textwidth]{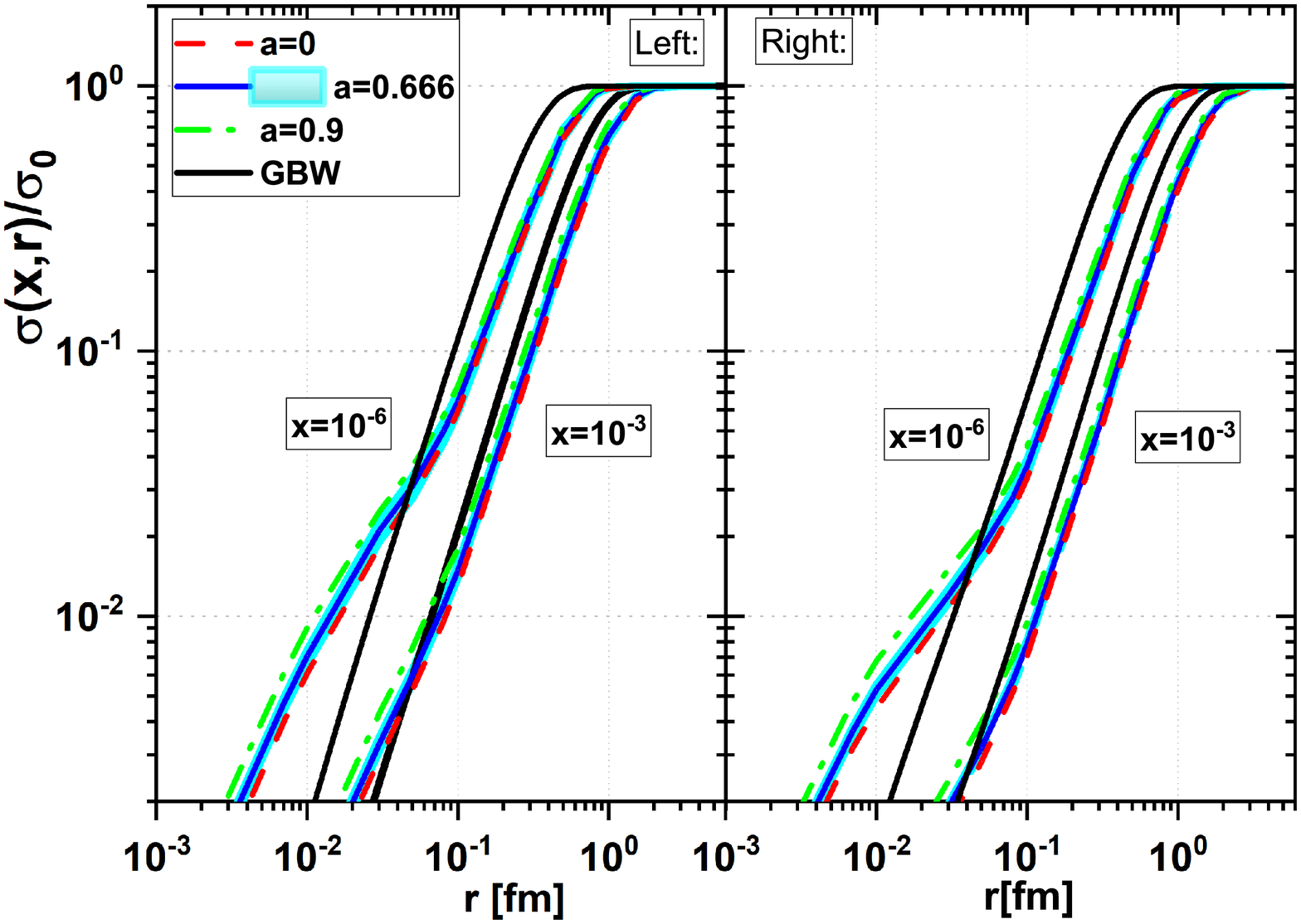}}
\caption{The ratio $\sigma_{\mathrm{dip}}(x,r)/\sigma_{0}$
according to the expansion points for $x=10^{-6}$ and $10^{-3}$
with the parameters compared with the GBW model: Left: Fit 1,1.
Right: Fit 2,2. The uncertainties in the expansion point $a=0.666$
are due to the statistical errors in Table I. }\label{Fig3}
\end{figure}
We see in the left and right plots of Fig.3 that for large values
of $r$ ($r{\gtrsim}1~\mathrm{fm}$) the two model results overlap
while the CDP results lie below the GBW curves for the interval
$0.04~\mathrm{fm}{\lesssim}r{\lesssim}1~\mathrm{fm}$. The
deviation of the CDP results from the GBW model in the left and
right plots of Fig.3 are visible for $r{\gtrsim}0.04~\mathrm{fm}$,
which is due to the running of the gluon distribution, is very
visible (especially at very low $x$). Expanding of the gluon
distribution around high and low values of $z=a$ is close to the
GBW results in the regions
$0.04~\mathrm{fm}{\lesssim}r{\lesssim}1~\mathrm{fm}$ and
$r{\lesssim}0.04~\mathrm{fm}$ respectively. The uncertainties, in
the left and right plots of Fig.3, are due to the statistical
errors in Table I in the expansion point $a=0.666$.\\
The form of the CDP results for the dipole cross-section
$\sigma_{\mathrm{dip}}(x,r)$ (i.e., Eq.(31)) at the expansion
points (a=0, 0.666 and 0.9) is shown in Fig.4 for $x=10^{-6}$ and
$10^{-2}$ with the parameters Fit 1,1 in Table I. In Fig.4, a
comparison with the GBW model (with charm contribution) is done in
a wide range of the dipole size $r$.  The CDP results also show
that the dipole cross-section features color transparency (i.e.,
$\sigma_{\mathrm{dip}}{\sim}r^2$) at small $r$ which is
perturbative QCD phenomenon and for large $r$, saturation occurs
(i.e., $\sigma_{\mathrm{dip}}{\simeq}\sigma_{0}$) [33]. We see
that the transition between two regimes occurs by decreasing
transverse sizes with a decrease of $x$ and this is visible in the
large
expansion points.\\
\begin{figure}
\centerline{
\includegraphics[width=0.58\textwidth]{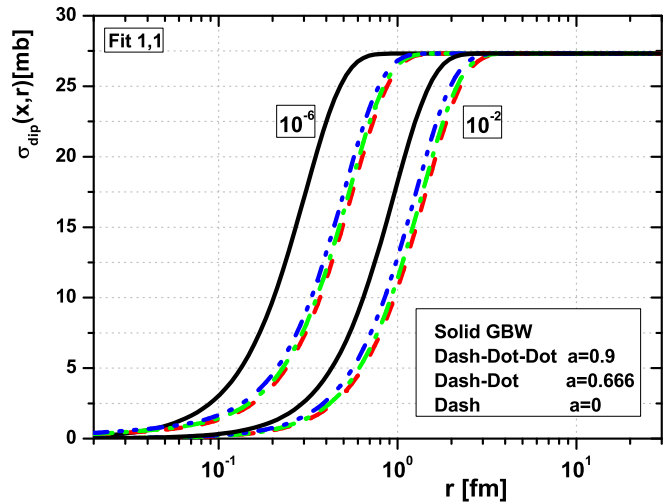}}
\caption{The dipole cross-section $\sigma_{\mathrm{dip}}(x,r)$ as
a function of transverse dipole size $r$  according to the
expansion points for $x=10^{-6}$ and $10^{-2}$ with the parameters
Fit 1,1 compared with the GBW model with charm
contribution.}\label{Fig3}
\end{figure}


\subsection{IV. Conclusions}

In summary, we have used an expansion method of the gluon density
in the color dipole formalism with heavy quark contributions and
obtain the dipole cross-section in a wide range of the transverse
dipole size $r$. The gluon density is obtained at an arbitrary
point of expansion of $G(\frac{x}{1-z})|_{z=a}$ where it$^{,}$s
dependent on the photoabsorption cross-section in the CDP. The
associated CDP gluon density is valid for large as well as small
$r$. The saturation line in the CDP model is dependent on the
running of the gluon density on the ($x, Q^2$)-plane and changed
than the GBW model at very low $x$.\\
The dipole cross-sections in the improved saturation model are
compared with the GBW model owing to the gluon density of the CDP
model. The dipole cross-section is modified at small $r$ than the
GBW model by the gluon density with the Altarelli-Martinelli
equation which is due to the pQCD phenomenon. The dipole
cross-section shows saturation features at large $r$ and this is
independent of the point of expansion of the gluon density, which
retains the features of the GBW model. The transition between the
saturation and color transparency regions is dependent on the $r$,
$x$ and the expansion point in our model. We can see that this
transition moves towards lower $r$ as $x$ decreases and the
expansion point increases.\\
In conclusion, the color dipole cross-section from unifying the
color dipole picture and improved saturation models gives a
behavior in accordance to the pQCD result with respect to the
expansion method.\\

\subsection{ACKNOWLEDGMENTS}
The author is grateful to Razi University for the financial
 support of this project.
I would like to thank D.Schildknecht and M.Kuroda for helpful
comments and discussions.\\

%

\section{References}
1. LHeC Collaboration, FCC-he Study Group, P. Agostini, et al., J.
Phys. G, Nucl. Part. Phys. {\bf48}, 110501 (2021).\\
2. FCC Collaboration, Eur.Phys.J.C {\bf79}, 474 (2019).\\
3. R.Abdul Khalek et al., Nucl. Phys.A {\bf1026}, 122447 (2022).\\
4. M.Klein, arXiv [hep-ph]:1802.04317.\\
5. D.Chakrabarti, P.Choudhary, B. Gurjat et al., Phys.Rev.D {\bf108}, 014009 (2023).\\
6. A.Bacchetta, F.G.Celiberto, M.Radici, and P.Taels, Eur.Phys.J.C
{\bf80}, 733 (2020).\\
7. A.V.Lipatov, G.I.Lykasov and M.A.Malyshev, Phys.Lett.B {\bf848},  138390 (2024).\\
8. Z. Lu and B.-Q. Ma, Phys.Rev.D {\bf94}, 094022 (2016).\\
9. N.Nikolaev and B.G.Zakharov, Z.Phys.C {\bf49}, 607 (1990).\\
10. K.Golec-Biernat and M.Wusthoff, Phys.Rev.D {\bf59}, 014017
(1998); Phys.Rev.D {\bf60}, 114023 (1999).\\
11. G.Cvetic, D.Schildknecht and A.Shoshi, Eur.Phys.J.C {\bf13},
301 (2000).\\
12. D.Schildknecht, Nucl.Phys.B (Proc. Suppl.) {\bf99}, 121
(2001); D.Schildknecht, B.Surrow and M.Tentyukov, Phys.Lett.B
{\bf499}, 116 (2001).\\
13. Yu S.Jeong, C.S.Kim, M. Vu Luu and M.H.Reno, JHEP {\bf11}, 025 (2014).\\
14. G.Cvetic, D.Schildknecht, B.Surrow and M.Tentyukov,
Eur.Phys.J.C {\bf 20}, 77 (2001).\\
15. M.Kuroda and D.Schildknecht, Phys.Rev.D {\bf85}, 094001
(2011); Int.J. of Mod.Phys.A {\bf31}, 1650157 (2016).\\
16. G.R.Boroun, M.Kuroda and D.Schildknecht, arXiv [hep-ph]: 2206.05672.\\
17. D.Schildknecht, Phys.Rev.D {\bf104}, 014009 (2021).\\
18. L.V.Gribov, E.M.Levin and M.G.Ryskin, Nucl.Phys.B {\bf188},
555 (1981) ; Phys.Rept. {\bf100},1 (1983).\\
19. A.H.Mueller and J.-w. Qiu, Nucl.Phys.B {\bf268}, 427 (1986).\\
20. I.Balitsky, Nucl.Phys.B {\bf463}, 99 (1996).\\
21. Y.V.Kovchegov, Phys.Rev.D {\bf60}, 034008 (1999) ; Phys.Rev.D
{\bf61}, 074018 (2000) 074018.\\
22. V.S.Fadin, E.A.Kuraev and L.N.Lipatov, Phys.Lett.B
\textbf{60}, 50(1975); L.N.Lipatov, Sov.J.Nucl.Phys. \textbf{23},
338(1976); I.I.Balitsky and L.N.Lipatov, Sov.J.Nucl.Phys.
\textbf{28}, 822(1978).\\
23. K.Kutak and A.M.Stasto, Eur.Phys.J.C {\bf41}, 343 (2005).\\
24. E.Iancu, A.Leonidov and L.McLerran, Nucl.Phys.A {\bf692}, 583
(2001); Phys.Lett.B {\bf510}, 133 (2001); E.Iancu,K.Itakura and
S.Munier, Phys.Lett.B {\bf590}, 199
(2004).\\
25. K. Golec-Biernat and S.Sapeta, JHEP {\bf03}, 102 (2018).\\
26. J.Bartels, K.Golec-Biernat and H.Kowalski, Phys. Rev.
D {\bf66}, 014001 (2002).\\
27. A.Luszczak, M.Luszczak and W.Schafer, arXiv:2210.02877.\\
28. G.Altarelli and G.Martinelli, Phys.Lett.B {\bf76}, 89
(1978).\\
29. S.Moch, J.A.M.Vermaseren and A.Vogt, Phys.Lett.B {\bf606}, 123
(2005).\\
30. M.B.Gay Ducati and P.B.Goncalves, Phys.Lett.B {\bf390}, 401
(1997).\\
31.  G.R.Boroun and B.Rezaei, Eur.Phys.J.C {\bf72}, 2221 (2012);
Phys.Rev.D {\bf105}, 034002 (2022); Phys.Letts.B {\bf816}, 136274
(2021); G.R.Boroun, Phys.Rev.C {\bf97}, 015206 (2018).\\
32. A.M.Cooper-Sarkar et.al., Z.Phys.C {\bf39}, 281(1988);
Acta.Phys.Polon.B {\bf34}, 2911(2003).\\
33. K. Golec-Biernat, J.Phys.G {\bf28}, 1057 (2002);
Acta.Phys.Polon.B {\bf33}, 2771 (2002).\\


\end{document}